# A double-hybrid density functional based on good local physics with outstanding performance on the GMTKN55 database

Axel D. Becke[1,a], Golokesh Santra[2], and Jan M. L. Martin[2]

[1]Department of Chemistry, Dalhousie University, 6274 Coburg Road, P.O. Box 15000, Halifax, Nova Scotia B3H 4R2, Canada

[2]Department of Molecular Chemistry and Materials Science, Weizmann Institute of Science, 7610001 Reḥovot, Israel

[a]Corresponding author: axel.becke@dal.ca

## Abstract

In two recent papers [A. D. Becke, J. Chem. Phys. **156**, 214101 (2022) and **157**, 234102 (2022)] we compared two Kohn-Sham density functionals based on *physical modelling* and *theory* with the best density-functional power-series fits in the literature. The best error statistics reported to date for a hybrid functional on the GMTKN55 chemical database of Goerigk, Grimme, and coworkers [Phys. Chem. Chem. Phys. **19**, 32184 (2017)] were obtained. In the present work, additional second-order perturbation-theory terms are considered. The result is a 12-parameter double-hybrid density functional with the lowest GMTKN55 "WTMAD2" error (1.76 kcal/mol) yet seen for any hybrid or double-hybrid density-functional approximation. We call it "DH23".

The GMTKN55 ("general main-group thermochemistry, kinetics, and noncovalent interactions") database of Goerigk, Grimme and coworkers[1] contains 1505 chemical reaction and barrier-height reference energies obtained from high-level wavefunction computations from over fifty sources. GMTKN55 and its predecessors, GMTKN24[2] and GMTKN30,[3] were compiled in order to assess the ever-growing number of density-functional theory (DFT) approximations in the literature. The database is composed of 55 subsets organized into five categories representing different chemical reaction types: basic properties, reactions between larger systems and isomerizations, barrier heights, and inter- and intra-molecular noncovalent interactions. Table S1 in the Supplementary Material describes the subsets in each category and Table S2 gives references to the original literature sources.

Although the intention of the GMTKN55 authors was the *assessment* of density-functional approximations (DFAs), its complexity and diversity make GMTKN55 highly attractive as a database for *fitting* of DFAs as well. The first (we believe) such fit was by Santra, Sylvetsky, and Martin,[4] who trained double-hybrid (DH) functionals on the full GMTKN55 data. Very recently, Becke[5,6] trained hybrid DFAs on a slightly reduced variant of GMTKN55. The focus of Becke's work was to test a suite of physically and theoretically motivated density-functional *models* against heavily fit functionals such as ωB97X-V[7] and ωB97M-V[8] from Mardirossian and Head-Gordon. Until Becke's work, ωB97X-V and ωB97M-V, trained on the massive MGCDB84[9] "main-group chemistry database," were the best-performing hybrid functionals tested on GMTKN55.[1,10] Since Santra and Martin had GMTKN55 second-order perturbation theory (PT2) correlation energies in hand from their DH computations, they suggested to Becke that it might be interesting to combine their PT2 energies with his DFT models. This Communication describes the outcome of that suggestion. In the end, we obtain a double-hybrid density functional with the lowest GMTKN55 error statistics reported to date.

To begin, we briefly review the exchange and correlation models used by Becke in his GMTKN55 trials. Full details can be found in Refs. 5 and 6. Our parent functional,[6] which will be appended by power-series and PT2 terms later, is

$$E_{XC}^0 = E_X^{exact} + a_{NDC}^{opp} U_{NDC}^{opp} + a_{NDC}^{par} U_{NDC}^{par} + a_{DC}^{opp} E_{DC}^{opp} + a_{DC}^{par} E_{DC}^{par} \tag{1}$$

$$+ a_X^{BR}(E_X^{BR} - E_X^{exact}) + E_{C6}^{XDM} + s_{RC}(E_{C8}^{XDM} + E_{C10}^{XDM}) \ .$$

The opposite-spin (opp) and parallel-spin (par) NDC terms are explicit models of *non-dynamical potential energy of correlation*[11] ("B05"). B05 is highly nonlocal, requiring the exact-exchange energy density at every grid point. The DC terms are total energies of dynamical correlation modelled as in Refs. 12 and 13. The term $a_X^{BR}(E_X^{BR} - E_X^{exact})$, involving the BR exchange model,[14] simulates non-dynamical correlation by virtue of the locality of BR exchange. The C6, C8, and C10 dispersion terms are from the XDM (exchange-hole dipole moment) model of Becke and Johnson.[15-18] The five "*a*" pre-factors and the $s_{RC}$ pre-factor allow fine tuning to GMTKN55 data by least-squares fitting. The "repulsion correction" factor $s_{RC}$ controls the interplay between the correlation terms and the dispersion terms to give correct Pauli repulsion when atomic densities overlap. All terms in the parent functional are fully described in the Appendices of Ref. 5.

Each term in Eq. (1) is based on an underlying reference-point-by-reference-point model of its associated exchange/correlation hole. Each model is governed by exact short-range behavior and normalization constraints at every point. The BR, NDC, and XDM models rely on the Taylor expansion of the exact spherically-averaged exchange hole to second order[19] as follows:

$$h_{X\sigma}^{exact}(\mathbf{r},s) = -\rho_\sigma - Q_\sigma s^2 + ... \qquad (2)$$

where

$$Q_\sigma = \frac{1}{6}\left(\nabla^2 \rho_\sigma - 2D_\sigma\right) \qquad (3)$$

and

$$D_\sigma = \tau_\sigma - \frac{1}{4}\frac{(\nabla \rho_\sigma)^2}{\rho_\sigma} \qquad (4)$$

and $\tau_\sigma$ is the following kinetic-energy density (without a ½ factor):

$$\tau_\sigma = \sum_i (\nabla \psi_{i\sigma})^2 \quad . \qquad (5)$$

In the above, $\mathbf{r}$ is the reference point, $s$ is the interelectronic distance, and $\sigma$ is the electron spin. $Q_\sigma$ is called the exchange-charge curvature. The DC dynamical correlation holes satisfy well-known interelectronic cusp conditions as discussed in Ref. 12. Thus, each term in Eq. (1) reflects "good local physics" (as asserted in the title) at every reference point.

In Ref. 6, exchange and correlation power series were added to Eq. (1) to investigate whether significant improvements could thereby be made. We invoked a dimensionless power-series variable

$$q_\sigma = \frac{Q_\sigma - Q_\sigma^{UEG}}{|Q_\sigma^{UEG}|} \tag{6}$$

where

$$Q_\sigma^{UEG} = -\frac{1}{5}(6\pi^2)^{2/3}\rho_\sigma^{5/3} \tag{7}$$

is the exchange-charge curvature of the uniform electron gas (UEG). This variable, too, reflects "good local physics". If $q_\sigma$ is positive (negative), then the local exchange-energy density is enhanced (attenuated) with respect to the UEG. Conversely, for dynamical correlation, positive (negative) $q_\sigma$ implies attenuation (enhancement) of the local correlation-energy density with respect to its UEG counterpart. $q_\sigma$ is predominantly positive in valence regions, and negative inside atomic cores.

As $q_\sigma$ has infinite range, $-\infty < q_\sigma < +\infty$, remapping into a finite domain is desirable. We use the same mapping for exchange, opposite-spin correlation, and parallel-spin correlation, but with different constants $\gamma_X$, $\gamma_{Copp}$, and $\gamma_{Cpar}$ for each:

$$w_{X\sigma} = \frac{\gamma_X q_\sigma}{\sqrt{1+\gamma_X^2 q_\sigma^2}} \quad , \quad w_{C\alpha\beta} = \frac{\gamma_{Copp} q_{\alpha\beta}}{\sqrt{1+\gamma_{Copp}^2 q_{\alpha\beta}^2}} \quad , \quad w_{C\sigma\sigma} = \frac{\gamma_{Cpar} q_\sigma}{\sqrt{1+\gamma_{Cpar}^2 q_\sigma^2}} \tag{8}$$

with

$$q_{\alpha\beta} = \frac{Q_\alpha + Q_\beta - Q_\alpha^{UEG} - Q_\beta^{UEG}}{|Q_\alpha^{UEG} + Q_\beta^{UEG}|} \tag{9}$$

and $\gamma_X = 0.11$, $\gamma_{Copp} = 0.14$, and $\gamma_{Cpar} = 1.6$. The range of the $w$ variable(s) is $-1 \le w \le +1$, and $w = 0$ corresponds to the UEG limit. We then add exchange, opposite-spin, and parallel-spin power series to the parent functional, Eq. (1), as follows:

$$E_{XC} = E_{XC}^0 + c_{X,0}(E_X^{UEG} - E_X^{exact}) + \sum_{i=1}^{m_X} c_{X,i} \sum_\sigma \int e_{X\sigma}^{UEG} w_{X\sigma}^i d^3\mathbf{r} \tag{10}$$

$$+ \sum_{i=0}^{m_{Copp}} c_{Copp,i} \int e_{C\alpha\beta}^{UEG} w_{C\alpha\beta}^i d^3\mathbf{r} + \sum_{i=0}^{m_{Cpar}} c_{Cpar,i} \sum_\sigma \int e_{C\sigma\sigma}^{UEG} f_\sigma^{SCC} w_{C\sigma\sigma}^i d^3\mathbf{r}$$

with respective expansion orders $m_X$, $m_{Copp}$, and $m_{Cpar}$. An additional *self-correlation correction*

$$f_\sigma^{SCC} = \frac{D_\sigma}{\tau_\sigma} \tag{11}$$

is inserted into the parallel-spin correlation terms to guarantee zero correlation energy in all one-electron systems. We subtract $E_X^{exact}$ from the $i = 0$ term in the exchange series to enforce the UEG limit. The UEG limit is not, however, enforced for correlation. Full details are available in Ref. 6, where this functional is called "B22plus".

Overall performance of a DFA on GMTKN55 will be assessed[1] by the Weighted Total Mean Absolute Deviation "WTMAD2":

$$\text{WTMAD2} = \frac{1}{\sum_{i=1}^{55} N_i} \sum_{i=1}^{55} N_i \frac{56.84 \, \text{kcal/mol}}{|\Delta E|_{\text{avg},i}} \text{MAD}_i \quad . \tag{12}$$

The MAD (mean absolute deviation) of each subset $i$ is weighted by the energy ratio 56.84/$|\Delta E|_{\text{avg},i}$ and the number of data $N_i$ in the set, as given in Table S1. The constant 56.84 kcal/mol is the average of $|\Delta E|_{\text{avg},i}$ over all the sets. Sixty-two of the 1505 reactions in GMTKN55 involve elements heavier than Kr: namely, all 28 reactions in the HEAVY28 set, 4 reactions in the HEAVYSB11 set, and 30 reactions in the HAL59 set. These require core pseudopotentials (PPs) in their computation. We omitted them in Refs. 5 and 6 in favor of the 1443 reactions involving all-electron calculations only. Here we do the same in our initial fits, to touch base with our earlier work. We call this "all-electron" database GMTKN54, and its analogous Eq. (12) error measure the WTMAD2(54). As the HEAVY28 set is excluded in the WTMAD2(54), the summations in Eq. (12) are over 54 sets instead of 55, but we retain the 56.84 kcal/mol constant and the $N_i$ values of Table S1 for compatibility with the literature. We will test on the full GMTKN55 database when our initial fits are completed.

Pre-factors and expansion coefficients are determined by least squares fitting. We fit to the GMTKN54 (or GMTKN55) data, taking the weight of each data point from Eq. (12) except for the W4-11 atomization energies, which are weighted by 1 (see Ref. 5). Note that this is *not* the same as minimizing the WTMAD2 directly, as has been done by the Martin group in Ref. 4 and later papers. For "B22plus" in Ref. 6, we chose optimum expansion orders $m_X = 2$, $m_{Copp} = 3$, and $m_{Cpar} = 4$, for a total of 12 series expansion coefficients. With the six pre-factors in $E_{XC}^0$ included, B22plus has 18 linear parameters altogether. We used BHandHLYP orbitals computed with the Gaussian 16 program[20] and the Dunning aug-cc-pVTZ basis set[21] (def2-TZVP[22] for K and Ca, and cc-pVTZ for the $C_{60}$ isomers). All correlation and power-series terms were computed using an in-house "postG" code. Technical details are given in the Supplementary Material, Appendix S1. The resulting B22plus WTMAD2(54) was 2.76 kcal/mol, substantially better than the previous best hybrid functionals, "B22" of Ref. 5 at 3.31 kcal/mol, and ωB97M-V at 3.37 kcal/mol. See Ref. 6 for additional information and statistics.

We now consider the further addition of second-order perturbation theory (PT2) terms in the *double*-hybrid DFA framework. Contemporary double hybrids originated with Grimme[23] and employ an MP2-like (Møller-Plesset perturbation theory) correlation component evaluated using Kohn-Sham DFT orbitals instead of Hartree-Fock (HF) orbitals. This is known as in Görling-Levy perturbation theory, GLPT2.[24] The benefits of using GLPT2 versus HF-MP2 as originally proposed in Ref. 25, are analyzed in Ref. 26. The authors of the GMTKN55 paper[1] clearly showed that double hybrids are superior to hybrid functionals which use only occupied orbitals. The literature on double hybrids is vast. A recent review by Martin and Santra[27] provides an overview of the current status of the field. Earlier reviews may be found in Refs. 28-30, and a comprehensive study of DH functionals on the GMTKN55 database has been presented by Mehta et al.[31]

ωB97M(2) from Mardirossian and Head-Gordon[32] is considered by many to be the leading double hybrid today. It has 14 independent parameters and is a "meta"-GGA (generalized gradient approximation) power series functional obtained after prescreening *trillions* of candidates. The WTMAD2 of ωB97M(2) on GMTKN55 is 2.13 kcal/mol.[33]

Santra, Semidalas, and Martin[34] have reported WTMAD2s as low as 1.85 kcal/mol by experimenting with an "XYG8" functional form that includes separate LDA (local density approximation) and GGA coefficients, and a sequence of input orbitals with varying amounts of exact-exchange fraction, as inspired by Refs. 35 and 36. These are not constrained to the correct asymptotic $C_6$ limit, as is ωB97M(2). They recover about 50-60% of the correct $C_6$.

In the present work, we add PT2 terms to Eqs. (1) and (10) in a manner that respects the correct asymptotic $C_6$ limit. Our functional therefore reflects not only good local physics, but good long-range physics as well. Independent opposite-spin and parallel-spin PT2 correlation energies are appended to Eq. (10) as follows:

$$E_{XC} = E_{XC}^0 + c_{X,0}(E_X^{UEG} - E_X^{exact}) + \sum_{i=1}^{m_X} c_{X,i} \sum_\sigma \int e_{X\sigma}^{UEG} w_{X\sigma}^i d^3\boldsymbol{r} \tag{13}$$

$$+ \sum_{i=0}^{m_{Copp}} c_{Copp,i} \int e_{C\alpha\beta}^{UEG} w_{C\alpha\beta}^i d^3\boldsymbol{r} + \sum_{i=0}^{m_{Cpar}} c_{Cpar,i} \sum_\sigma \int e_{C\sigma\sigma}^{UEG} f_\sigma^{SCC} w_{C\sigma\sigma}^i d^3\boldsymbol{r}$$

$$+ a_{PT2}^{opp}(E_{PT2}^{opp} - \frac{1}{2}E_{C6}^{XDM}) + a_{PT2}^{par}(E_{PT2}^{par} - \frac{1}{2}E_{C6}^{XDM})$$

where

$$E_{PT2}^{par} = E_{PT2}^{\alpha\alpha} + E_{PT2}^{\beta\beta} \quad . \tag{14}$$

The logic in the added PT2 terms is simple. Lochan, Jung, and Head-Gordon[37] have stated that, at long range for two non-overlapping closed-shell systems, the opposite-spin and parallel-spin parts of the inter-system MP2 correlation energy are equal. Assuming that MP2 (PT2) and XDM give the correct asymptotic dispersion energy, we therefore have

$$E_{PT2}^{opp} = E_{PT2}^{par} = \frac{1}{2} E_{C6}^{XDM} \qquad (15)$$

at long range. The parent functional, Eq. (1), contains a full $E_{C6}^{XDM}$ term. Hence the PT2 terms in Eq. (13) must vanish asymptotically. By Eq. (15), they do.

PT2 energies are computed by the Q-Chem 6.0 program[38] using "resolution of the identity" (RI-MP2) speed-ups. As is common practice in DH applications, single-excitation terms, non-zero in principle in GLPT2, have been omitted. For the PT2 correlation energies, the same frozen-core settings were used as in Ref. 4. These calculations were run on the Weizmann Faculty of Chemistry's HPC facility "ChemFarm".

In our preliminary explorations, we discovered some welcome simplifications of Eq. (13). Power-series orders $m_X$, $m_{Copp}$, and $m_{Cpar}$ beyond quadratic offered no improvements. Furthermore, the entire exchange series can be removed with negligible degradation of the WTMAD2(54). Most significantly, the B05 non-dynamical correlation terms in the parent functional, Eq. (1), are *not important* in a DH context despite being critical to the success of B22plus.[6] PT2 correlation handily compensates B05, with WTMAD2(54) affected by less than ~ 0.05 kcal/mol if the latter is omitted. This is a major win, as the exact-exchange energy density required by B05 at every grid point is available in very few quantum chemistry programs. Hereafter, then, all results pertain to the simplified functional

$$E_{XC} = E_X^{exact} + a_{DC}^{opp} E_{DC}^{opp} + a_{DC}^{par} E_{DC}^{par} + a_X^{BR}(E_X^{BR} - E_X^{exact}) \quad (16)$$

$$+ E_{C6}^{XDM} + s_{RC}(E_{C8}^{XDM} + E_{C10}^{XDM})$$

$$+ \sum_{i=0}^{2} c_{Copp,i} \int e_{C\alpha\beta}^{UEG} w_{C\alpha\beta}^i d^3\mathbf{r} + \sum_{i=0}^{2} c_{Cpar,i} \sum_{\sigma} \int e_{C\sigma\sigma}^{UEG} f_{\sigma}^{SCC} w_{C\sigma\sigma}^i d^3\mathbf{r}$$

$$+ a_{PT2}^{opp}(E_{PT2}^{opp} - \frac{1}{2}E_{C6}^{XDM}) + a_{PT2}^{par}(E_{PT2}^{par} - \frac{1}{2}E_{C6}^{XDM})$$

with 12 linear coefficients altogether. This is two fewer than the number of parameters in ωB97M(2). We also note that, after eliminating the exchange series from Eq. (13), Eq. (16) *exactly* models the hydrogen atom at *every* reference point. Only BR-based functionals have this nice property.[14] We shall call the functional of Eq. (16) "DH23".

To connect with our previous work,[5,6] first tests have been carried out with BHandHLYP orbitals and the Dunning aug-cc-pVTZ basis set[21] (def2-TZVP[22] for K and Ca, and cc-pVTZ for the $C_{60}$ isomers). The resulting WTMAD2(54) is 2.18 kcal/mol, about 0.6 kcal/mol lower than for our B22plus hybrid. Next, we expand from the Dunning triple-zeta basis to the quadruple-zeta def2-QZVPP basis.[22] For subsets involving anions (AHB21, BH76 and BH76RC, G21EA, IL16, WATER27) or rare-gas interactions (RG18), diffuse functions are added using def2-QZVPPD.[39] We drop down, however, to def2-TZVPP for the very large UPU23 and C60ISO systems and the two largest isomers (1 and 4) in the ISOL24 subset. The BHandHLYP quadruple-zeta

WTMAD2(54) is 1.76 kcal/mol, a substantial reduction from the Dunning triple-zeta 2.18 kcal/mol value.

BHandHLYP orbitals incorporate 50% Hartree-Fock (HF) exchange. It has been argued by Goerigk and Grimme[3] and there is systematic computational evidence[34] that orbitals with 20-30% HF may offer better DH performance. We explore this by using "B1LYP" orbitals[40] with 25% HF exchange. We observe a slight decrease in the WTMAD2(54) from 1.76 (BHandHLYP) to 1.72 (B1LYP) kcal/mol, employing the same def2 basis sets described above. For completeness, B1LYP calculations have also been performed using diffuse basis functions everywhere (def2-QZVPPD, except def2-TZVPPD for UPU23, C60ISO, and the two largest ISOL24 isomers, 1 and 4). Universal diffuse functions have a minor effect on the WTMAD2(54), giving a value of 1.74 kcal/mol. Similar observations were made in Ref. 41.

The 1.72 kcal/mol WTMAD2(54) achieved here is more than 1 kcal/mol lower than the 2.76 kcal/mol of B22plus, our best hybrid functional of Ref. 6, and does not depend on the cumbersome B05 non-dynamical correlation energy. Readily available and economical (thanks to resolution of the identity RI-MP2 technology[42-44]) PT2 correlation is used instead. How important are the power-series terms in Eq. (16)? Omitting them loses six parameters, but raises our best B1LYP def2-QZVPP WTMAD2(54) from 1.72 kcal/mol to 2.11 kcal/mol, almost 0.4 kcal/mol higher. This is nevertheless rather good for a DH with only six parameters. Table I summarizes our WTMAD2(54) results for BHandHLYP vs. B1LYP input orbitals, and the aug-cc-pVTZ vs. def2-QZVPP basis sets.

Having established, on a foundation of all-electron calculations, that B1LYP orbitals and the def2-QZVPP basis set are good choices, we now perform a fit on the full GMTKN55 database including the 62 reactions whose calculations require core pseudopotentials. To accommodate PPs, additional considerations and modifications beyond the descriptions in Refs. 5 and 6 were necessary for our XDM dispersion model. We defer discussion of these to the Supplementary Material, Appendix S2.

The GMTKN55 fit coefficients are given in Table II. The exact exchange fraction is a sizable 83.6%. The UEG limit for opposite-spin correlation is 0.86, not far from 1, and for parallel-spin correlation is roughly 0.92 (using that the UEG limit of $E_{DC}^{par}$ is about half the correct value[12]). The negative value, -0.299, of the repulsion-correction scale factor $s_{RC}$ implies that PT2 overestimates dispersion energy at intermediate range. This echoes previous findings of the Martin group.[27,45]

In Table III, we list WTMAD2s for each of the GMTKN55 chemical categories and for the complete database. For comparison, we include data for four additional leading DH functionals: ωB97M(2)[32] from Mardirossian and Head-Gordon, and XYG8[*f1*]@B$_{20}$LYP,[34] revDSD-PBEP86-D4,[33] and xDSD$_{75}$-PBEP86-D4[33] from the Martin group. The number of parameters in these functionals is 14 for ωB97M(2) and 8 for the last three, versus 12 for DH23. All computations employ the def2-QZVPP basis set with exceptions as noted earlier. DH23 has the smallest WTMAD2 on the full GMTKN55 database at 1.76 kcal/mol, followed by XYG8[*f1*]@B$_{20}$LYP at 1.85 kcal/mol. The categories "basic properties of small systems" and "reaction and isomerization energies of large systems" are clearly led by DH23, but XYG8[*f1*]@B$_{20}$LYP has a slight edge for barrier heights and (all) noncovalent interactions. See Table S3 in the Supplementary Material for MADs of all the considered functionals on all the GMTKN55 subsets. Noteworthy is the superior performance of DH23 on the "mindless benchmarking" MB16-43 set of artificial molecules, designed to test the robustness of DFAs. Table III is represented by a radar plot in Fig. 1.

Let us comment, finally, on the asymptotic $C_6$ limit. DH23 and ωB97M(2) recover the correct limit by design. XYG8[*f1*]@B$_{20}$LYP, revDSD-PBEP86-D4, and xDSD$_{75}$-PBEP86-D4 recover about 50%, 90%, and 80% respectively. The importance of enforcing the correct $C_6$ limit is debatable in molecules near equilibrium structures. See Wu and Truhlar[46] for an assessment on very large systems containing 200 to 910 atoms. Nevertheless, good physics is our goal for DH23.

Of the many hundreds of hybrid and double-hybrid DFAs tested on the GMTKN55 database so far, DH23, at 1.76 kcal/mol, has the lowest WTMAD2. This is gratifying given the simple manner in which it was derived. ωB97M(2) was derived from trillions of prescreening fits. The functionals XYG8[*f1*]@B$_{20}$LYP, revDSD-PBEP86-D4, and xDSD$_{75}$-PBEP86-D4 are the result of many trials with the zoo[1] of known local-density, GGA, and meta-GGA exchange and correlation functionals. DH23, on the other hand, is based on explicit exchange-correlation hole models satisfying exact short-range-behavior and normalization constraints at every point, and a power-series variable $q_\sigma$ governing the enhancement/attenuation of exchange and correlation energy densities with respect to the UEG, again, at every point. With minimal experimentation, DH23 was quickly and straightforwardly derived from its hybrid-functional parents.[5,6]

The calibrations of this work are on main-group and organic thermochemistry, including noncovalent interactions. Preliminary tests on 3d transition-metal reactions, described in Appendix S3 of the Supplementary Material, are encouraging. A full assessment of the performance of DH23 in transition-metal chemistry will be undertaken in future work.

## Supplementary Material

See the Supplementary Material for tables of the 55 subsets of the GMTKN55 database and their original literature references, and the GMTKN55 MADs for the five DH functionals compared in this work. Also, technical details of the "postG" calculations, modifications of the XDM dispersion formulas to accommodate core pseudopotentials, and preliminary 3d transition-metal results, are presented in Appendices S1 to S3.

## Acknowledgments

A.D.B. gratefully acknowledges the support of the Natural Sciences and Engineering Research Council of Canada (NSERC) and Prof. Erin Johnson (Dalhousie University) for providing computing access through the Digital Research Alliance of Canada. G.S. acknowledges a doctoral fellowship from the Feinberg Graduate School (Weizmann Institute of Science). Research at Weizmann was supported by the Israel Science Foundation (grant 1969/20), by the Minerva Foundation (grant 2020/05), and by a research grant from the Dr. Uriel Arnon Memorial Fund for Artificial Intelligence and Smart Materials.

Table I: WTMAD2(54) on GMTKN54, in kcal/mol.

| Input orbitals/basis set | Eq. (16) | Eq. (16) without power series |
|---|---|---|
| BHandHLYP/aVTZ[a] | 2.18 | 2.24 |
| BHandHLYP/QZVPP[b] | 1.76 | 2.05 |
| B1LYP/QZVPP[b] | 1.72 | 2.11 |
| B1LYP/QZVPPD[c] | 1.74 | 2.14 |
|  | (12 parameters) | (6 parameters) |

a) aug-cc-pVTZ with exceptions noted in text.

b) def2-QZVPP with exceptions noted in text.

c) def2-QZVPPD with exceptions noted in text.

Table II: Eq. (16) coefficients for B1LYP orbitals and QZVPP[a] basis set, fit to the full GMTKN55 database.

| | | | | | |
|---|---|---|---|---|---|
| $a_{DC}^{opp}$ | 0.010 | $c_{Copp,0}$ | 0.449 | $a_{PT2}^{opp}$ | 0.403 |
| $a_{DC}^{par}$ | 1.536 | $c_{Copp,1}$ | -0.240 | $a_{PT2}^{par}$ | 0.193 |
| $a_X^{BR}$ | 0.164 | $c_{Copp,2}$ | 0.553 | | |
| $s_{RC}$ | -0.299 | $c_{Cpar,0}$ | -0.040 | | |
| | | $c_{Cpar,1}$ | -0.487 | | |
| | | $c_{Cpar,2}$ | 0.772 | | |

a) def2-QZVPP with exceptions noted in text.

Table III: WTMAD2 (kcal/mol) for GMTKN55 categories.[a]

| DH23[b] | ωB97M(2)[c] | XYG8[d] | revDSD[e] | xDSD[f] |
|---|---|---|---|---|

*Basic properties and reaction energies of small systems:*

| 1.18 | 1.39 | 1.42 | 1.75 | 1.63 |
|---|---|---|---|---|

*Reaction energies for large systems and isomerization reactions:*

| 2.06 | 2.61 | 2.32 | 3.59 | 3.07 |
|---|---|---|---|---|

*Reaction barrier heights:*

| 1.81 | 1.66 | 1.65 | 2.02 | 1.95 |
|---|---|---|---|---|

*Intermolecular noncovalent interactions:*

| 1.99 | 2.44 | 1.99 | 2.29 | 2.28 |
|---|---|---|---|---|

*Intramolecular noncovalent interactions:*

| 2.19 | 3.00 | 2.17 | 2.08 | 2.11 |
|---|---|---|---|---|

*All noncovalent interactions:*

| 2.09 | 2.72 | 2.07 | 2.18 | 2.20 |
|---|---|---|---|---|

*All GMTKN55:*

| 1.76 | 2.13 | 1.85 | 2.25 | 2.12 |
|---|---|---|---|---|

a) See Table S1 in the Supplementary Material.

b) Present work, Eq. (16).

c) Ref. 32.

d) XYG8[*f1*]@B$_{20}$LYP, Ref. 34 (parameter values in line 5 of Table 1).

e) revDSD-PBEP86-D4, Ref. 33 (parameter values in line 6 of Table S2).

f) xDSD$_{75}$-PBEP86-D4, Ref. 33 (parameter values in line 1 of Table 1).

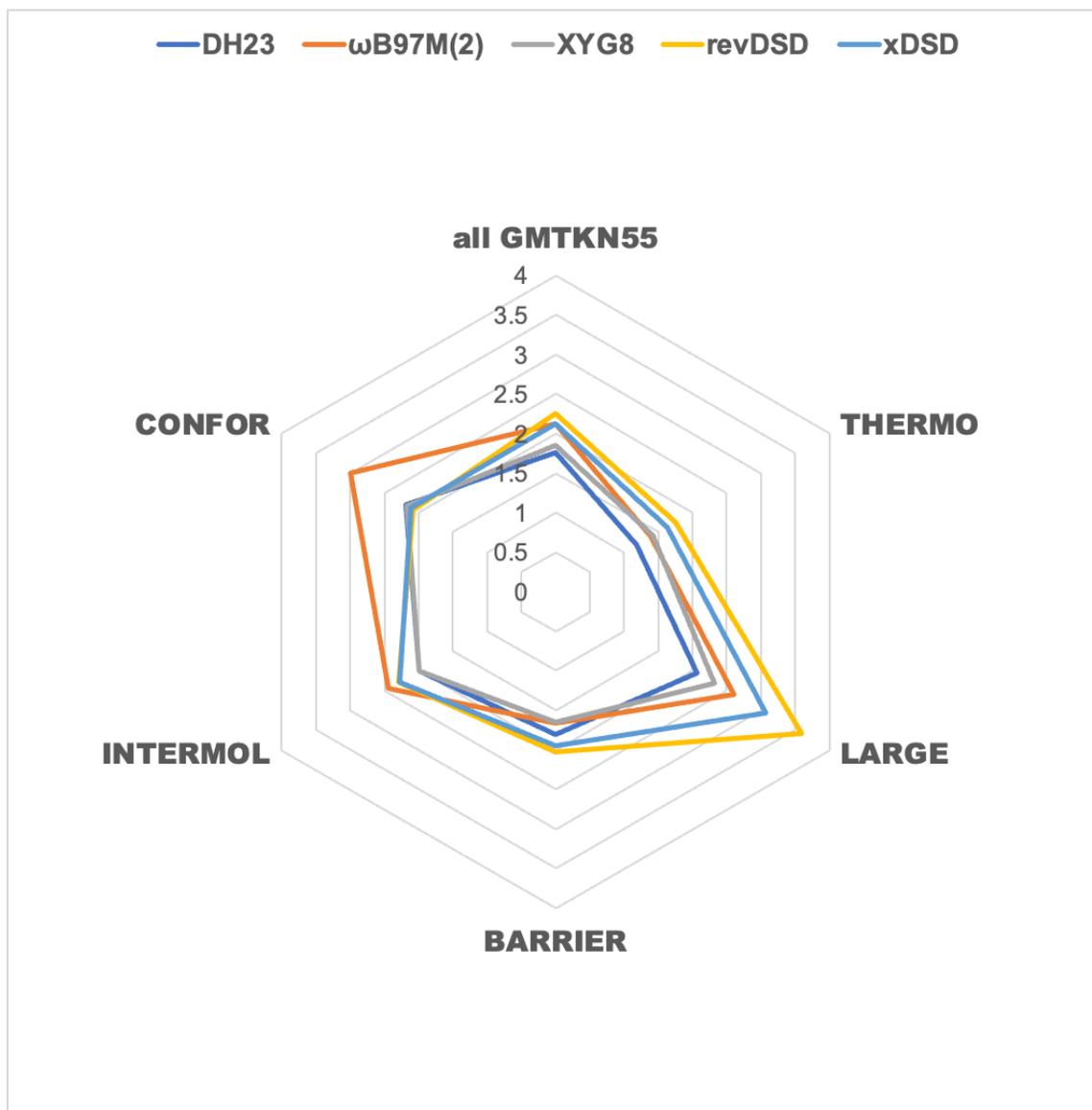

Figure1: Radar plot of WTMAD2s of GMTKN55 chemical categories in Table III. Functional designations as in the Table III header.

THERMO: *Basic properties and reaction energies of small systems.* LARGE: *Reaction energies for large systems and isomerization reactions.* BARRIER: *Reaction barrier heights.* INTERMOL: *Intermolecular noncovalent interactions.* CONFOR: *Intramolecular noncovalent interactions.*

Supplementary Material

A double-hybrid density functional based on good local physics with outstanding performance on the GMTKN55 database

Axel D. Becke[1,a], Golokesh Santra[2], and Jan M. L. Martin[2]

[1]Department of Chemistry, Dalhousie University, 6274 Coburg Road, P.O. Box 15000, Halifax, Nova Scotia B3H 4R2, Canada

[2]Department of Molecular Chemistry and Materials Science, Weizmann Institute of Science, 7610001 Reḥovot, Israel

a)Corresponding author: axel.becke@dal.ca

# Table S1: GMTKN55 database subsets.

| | | $N^a$ | $|\Delta E|_{avg}^b$ |
|---|---|---|---|
| Basic properties and reaction energies of small systems: | | | |
| W4-11 | Total atomization energies | 140 | 306.91 |
| G21EA | Adiabatic electron affinities | 25 | 33.62 |
| G21IP | Adiabatic ionization potentials | 36 | 257.61 |
| DIPCS10 | Double-ionization potentials of closed-shell systems | 10 | 654.26 |
| PA26 | Adiabatic proton affinities (incl. of amino acids) | 26 | 189.05 |
| SIE4x4 | Self-interaction-error related problems | 16 | 33.72 |
| ALKBDE10 | Dissociation energies of group-1 and -2 diatomics | 10 | 100.69 |
| YBDE18 | Bond-dissociation energies in ylides | 18 | 49.28 |
| AL2x6 | Dimerization energies of $AlX_3$ compounds | 6 | 35.88 |
| HEAVYSB11 | Dissociation energies of heavy-element compounds | 11 | 58.02 |
| NBPRC | Oligomerizations and $H_2$ fragmentations of $NH_3/BH_3$ systems | 12 | 27.71 |
| | $H_2$ activation reactions with $PH_3/BH_3$ systems | | |
| ALK8 | Dissociation and other reactions of alkaline compounds | 8 | 62.60 |
| RC21 | Fragmentations and rearrangements in radical cations | 21 | 35.70 |
| G2RC | Reaction energies of selected G2/97 systems | 25 | 51.26 |
| BH76RC | Reaction energies of the BH76 set | 30 | 21.39 |
| FH51 | Reaction energies in various (in-)organic systems | 51 | 31.01 |
| TAUT15 | Relative energies of tautomers | 15 | 3.05 |
| DC13 | 13 difficult cases for DFT methods | 13 | 54.98 |
| Reaction energies for large systems and isomerization reactions: | | | |
| MB16-43 | Decomposition energies of artificial molecules | 43 | 414.73 |
| DARC | Reaction energies of Diels–Alder reactions | 14 | 32.47 |
| RSE43 | Radical-stabilization energies | 43 | 7.60 |
| BSR36 | Bond-separation reactions of saturated hydrocarbons | 36 | 16.20 |
| CDIE20 | Double-bond isomerization energies in cyclic systems | 20 | 4.06 |
| ISO34 | Isomerization energies of small and medium-sized organic molecules | 34 | 14.57 |
| ISOL24 | Isomerization energies of large organic molecules | 24 | 21.92 |
| C60ISO | Relative energies of $C_{60}$ isomers | 9 | 98.25 |
| PArel | Relative energies of protonated isomers | 20 | 4.63 |

Reaction barrier heights:

| | | | |
|---|---|---|---|
| BH76 | Barrier heights of hydrogen transfer, heavy atom transfer, nucleophilic substitution, unimolecular and association reactions | 76 | 18.61 |
| BHPERI | Barrier heights of pericyclic reactions | 26 | 20.87 |
| BHDIV10 | Diverse reaction barrier heights | 10 | 45.33 |
| INV24 | Inversion/racemization barrier heights | 24 | 31.85 |
| BHROT27 | Barrier heights for rotation around single bonds | 27 | 6.27 |
| PX13 | Proton-exchange barriers in $H_2O$, $NH_3$, and HF clusters | 13 | 33.36 |
| WCPT18 | Proton-transfer barriers in uncatalyzed and water-catalyzed reactions | 18 | 34.99 |

Intermolecular noncovalent interactions:

| | | | |
|---|---|---|---|
| RG18 | Interaction energies of rare-gas complexes | 18 | 0.58 |
| ADIM6 | Interaction energies of n-alkane dimers | 6 | 3.36 |
| S22 | Binding energies of noncovalently bound dimers | 22 | 7.30 |
| S66 | Binding energies of noncovalently bound dimers | 66 | 5.47 |
| HEAVY28 | Noncovalent interaction energies between heavy element hydrides | 28 | 1.24 |
| WATER27 | Binding energies in $(H_2O)_n$, $H^+(H_2O)_n$ and $OH^-(H_2O)_n$ | 27 | 81.14 |
| CARBHB12 | Hydrogen-bonded complexes between carbene analogs and $H_2O$, $NH_3$, or HCl | 12 | 6.04 |
| PNICO23 | Interaction energies in pnicogen-containing dimers | 23 | 4.27 |
| HAL59 | Binding energies in halogenated dimers (incl. halogen bonds) | 59 | 4.59 |
| AHB21 | Interaction energies in anion–neutral dimers | 21 | 22.49 |
| CHB6 | Interaction energies in cation–neutral dimers | 6 | 26.79 |
| IL16 | Interaction energies in anion–cation dimers | 16 | 109.04 |

Intramolecular noncovalent interactions:

| | | | |
|---|---|---|---|
| IDISP | Intramolecular dispersion interactions | 6 | 14.22 |
| ICONF | Relative energies of conformers of inorganic systems | 17 | 3.27 |
| ACONF | Relative energies of alkane conformers | 15 | 1.83 |
| AMINO20x4 | Relative energies of amino acid conformers | 80 | 2.44 |
| PCONF21 | Relative energies of tri- and tetrapeptide conformers | 18 | 1.62 |
| MCONF | Relative energies of melatonin conformers | 51 | 4.97 |
| SCONF | Relative energies of sugar conformers | 17 | 4.60 |
| UPU23 | Relative energies between RNA-backbone conformers | 23 | 5.72 |
| BUT14DIOL | Relative energies of butane-1,4-diol conformers | 64 | 2.80 |

a) Number of reactions in subset.
b) Averaged absolute reaction energy (kcal/mol).

## Table S2: GMTKN55 database references.

**ACONF**[1] Relative energies of alkane conformers

**ADIM6**[2] Interaction energies of n-alkane dimers

**AHB21**[3] Interaction energies in anion–neutral dimers

**AL2X6**[4] Dimerization energies of $AlX_3$ compounds

**ALK8**[4] Dissociation and other reactions of alkaline compounds

**ALKBDE10**[5] Dissociation energies in group-1 and -2 diatomics

**AMINO20x4**[6] Relative energies in amino acid conformers

**BH76RC**[7] Reaction energies of the BH76[7-9] set

**BH76**[7-9] Barrier heights of hydrogen transfer, heavy atom transfer, nucleophilic substitution, unimolecular and association reactions

**BHDIV10**[4] Diverse reaction barrier heights

**BHPERI**[4,10-12] Barrier heights of pericyclic reactions

**BHROT27**[4] Barrier heights for rotation around single bonds

**BSR36**[13,14] Bond-separation reactions of saturated hydrocarbons

**BUT14DIOL**[15] Relative energies in butane-1,4-diol conformers

**C60ISO**[16] Relative energies between $C_{60}$ isomers

**CARBHB12**[4] Hydrogen-bonded complexes between carbene analogues and $H_2O$, $NH_3$, or HCl

**CDIE20**[17] Double-bond isomerization energies in cyclic systems

**CHB6**[3] Interaction energies in cation–neutral dimers

**DARC**[7,18] Reaction energies of Diels–Alder reactions

**DC13**[19,7,20–29] 13 difficult cases for DFT methods

**DIPCS10**[4] Double-ionization potentials of closed-shell systems

**FH51**[30,31] Reaction energies in various (in-)organic systems

**G21EA**[7,32] Adiabatic electron affinities

**G21IP**[7,32] Adiabatic ionization potentials

**G2RC**[1,33] Reaction energies of selected G2/97 systems

**HAL59**[34,35] Binding energies in halogenated dimers (incl. halogen bonds)

**HEAVY28**[9] Noncovalent interaction energies between heavy element hydrides

**HEAVYSB11**[4] Dissociation energies in heavy-element compounds

**ICONF**[4] Relative energies in conformers of inorganic systems

**IDISP**[7,36–39] Intramolecular dispersion interactions

**IL16**[3] Interaction energies in anion–cation dimers

**INV24**[40] Inversion/racemization barrier heights

**ISO34**[36] Isomerization energies of small and medium-sized organic molecules

**ISOL24**[41] Isomerization energies of large organic molecules

**MB16-43**[4] Decomposition energies of artificial molecules

**MCONF**[42] Relative energies in melatonin conformers

**NBPRC**[7,38,43] Oligomerizations and $H_2$ fragmentations of $NH_3/BH_3$ systems; $H_2$ activation reactions with $PH_3/BH_3$ systems

**PA26**[4] Adiabatic proton affinities (incl. of amino acids)

**PArel**[4] Relative energies in protonated isomers

**PCONF21**[4] Relative energies in tri- and tetrapeptide conformers

**PNICO23**[44] Interaction energies in pnicogen-containing dimers

**PX13**[45] Proton-exchange barriers in $H_2O$, $NH_3$, and HF clusters

**RC21**[4] Fragmentations and rearrangements in radical cations

**RG18**[4] Interaction energies in rare-gas complexes

**RSE43**[46] Radical-stabilization energies

**S22**[47] Binding energies of noncovalently bound dimers

**S66**[48] Binding energies of noncovalently bound dimers

**SCONF**[7,49] Relative energies of sugar conformers

**SIE4x4**[50] Self-interaction-error related problems

**TAUT15**[4] Relative energies in tautomers

**UPU23**[51] Relative energies between RNA-backbone conformers

**W4-11**[52] Total atomization energies

**WATER27**[53] Binding energies in $(H_2O)_n$, $H^+(H_2O)_n$ and $OH^-(H_2O)_n$

**WCPT18**[54] Proton-transfer barriers in uncatalyzed and water-catalyzed reactions

**YBDE18**[55] Bond-dissociation energies in ylides

Table S3: MADs (kcal/mol) of DH functionals on GMTKN55 subsets.

| | DH23[a] | ωB97M(2)[b] | XYG8[c] | revDSD[d] | xDSD[e] |
|---|---|---|---|---|---|
| W4-11 | 2.03 | 1.38 | 2.30 | 2.49 | 2.07 |
| G21EA | 1.43 | 1.50 | 1.91 | 2.57 | 2.70 |
| G21IP | 1.19 | 1.60 | 1.77 | 2.27 | 2.12 |
| DIPCS10 | 1.15 | 2.53 | 4.48 | 4.47 | 4.78 |
| PA26 | 0.69 | 1.06 | 0.85 | 1.37 | 1.29 |
| SIE4x4 | 1.81 | 5.25 | 1.38 | 4.94 | 3.85 |
| ALKBDE10 | 3.55 | 3.54 | 4.88 | 2.95 | 2.80 |
| YBDE18 | 0.96 | 0.98 | 0.60 | 0.73 | 0.84 |
| AL2X6 | 0.42 | 1.41 | 0.49 | 1.69 | 1.41 |
| HEAVYSB11 | 1.00 | 0.84 | 1.02 | 1.46 | 1.11 |
| NBPRC | 0.88 | 0.73 | 0.61 | 0.73 | 0.60 |
| ALK8 | 0.74 | 2.12 | 1.13 | 2.08 | 1.83 |
| RC21 | 0.64 | 1.08 | 0.82 | 1.50 | 1.77 |
| G2RC | 0.84 | 1.36 | 1.33 | 1.51 | 1.34 |
| BH76RC | 0.94 | 0.81 | 0.98 | 0.79 | 0.72 |
| FH51 | 0.68 | 0.65 | 0.72 | 0.72 | 0.65 |
| TAUT15 | 0.30 | 0.27 | 0.48 | 0.53 | 0.51 |
| DC13 | 2.42 | 1.59 | 3.54 | 1.74 | 1.85 |
| MB16-43 | 4.35 | 13.62 | 11.62 | 14.37 | 10.46 |
| DARC | 0.50 | 1.56 | 0.55 | 0.49 | 0.37 |
| RSE43 | 0.21 | 0.44 | 0.15 | 0.83 | 0.76 |
| BSR36 | 0.91 | 0.64 | 0.70 | 1.34 | 0.77 |
| CDIE20 | 0.15 | 0.22 | 0.23 | 0.35 | 0.33 |
| ISO34 | 0.30 | 0.38 | 0.42 | 0.37 | 0.38 |
| ISOL24 | 0.86 | 1.02 | 1.30 | 1.13 | 1.02 |
| C60ISO | 9.58 | 2.44 | 9.94 | 6.24 | 7.47 |
| PArel | 0.39 | 0.43 | 0.38 | 0.36 | 0.36 |
| BH76 | 0.78 | 0.65 | 0.68 | 0.93 | 0.95 |
| BHPERI | 0.39 | 0.51 | 0.42 | 0.75 | 0.72 |
| BHDIV10 | 0.70 | 0.77 | 0.93 | 0.80 | 0.66 |
| INV24 | 0.55 | 0.94 | 0.58 | 0.67 | 0.70 |
| BHROT27 | 0.14 | 0.17 | 0.16 | 0.10 | 0.09 |
| PX13 | 2.04 | 0.55 | 0.95 | 1.50 | 1.17 |
| WCPT18 | 1.02 | 1.07 | 1.21 | 0.89 | 0.73 |
| RG18 | 0.05 | 0.07 | 0.05 | 0.08 | 0.08 |
| ADIM6 | 0.32 | 0.43 | 0.34 | 0.26 | 0.36 |
| S22 | 0.16 | 0.14 | 0.12 | 0.14 | 0.11 |
| S66 | 0.16 | 0.15 | 0.16 | 0.16 | 0.17 |
| HEAVY28 | 0.08 | 0.10 | 0.06 | 0.10 | 0.09 |
| WATER27 | 0.65 | 0.37 | 1.45 | 0.53 | 0.69 |
| CARBHB12 | 0.22 | 0.22 | 0.25 | 0.36 | 0.31 |
| PNICO23 | 0.08 | 0.21 | 0.06 | 0.10 | 0.08 |
| HAL59 | 0.20 | 0.27 | 0.22 | 0.22 | 0.23 |
| AHB21 | 0.37 | 0.25 | 0.35 | 0.23 | 0.24 |
| CHB6 | 0.88 | 0.87 | 1.14 | 0.70 | 0.72 |
| IL16 | 0.64 | 0.26 | 0.70 | 0.36 | 0.24 |
| IDISP | 0.43 | 1.29 | 0.33 | 0.58 | 0.66 |
| ICONF | 0.09 | 0.12 | 0.12 | 0.11 | 0.11 |
| ACONF | 0.07 | 0.11 | 0.08 | 0.03 | 0.05 |
| Amino20x4 | 0.08 | 0.08 | 0.09 | 0.12 | 0.12 |
| PCONF21 | 0.16 | 0.33 | 0.17 | 0.12 | 0.11 |
| MCONF | 0.24 | 0.32 | 0.26 | 0.09 | 0.09 |
| SCONF | 0.05 | 0.22 | 0.04 | 0.07 | 0.06 |
| UPU23 | 0.39 | 0.44 | 0.39 | 0.49 | 0.47 |
| BUT14DIOL | 0.06 | 0.05 | 0.02 | 0.05 | 0.06 |

a) Present work, Eq. (16) in the text.

b) Ref. 32 in text.

c) XYG8[$f1$]@B$_{20}$LYP, Ref. 34 in text (parameter values in line 5 of Table 1).

d) revDSD-PBEP86-D4, Ref. 33 in text (parameter values in line 6 of Table S2).

e) xDSD$_{75}$-PBEP86-D4, Ref. 33 in text (parameter values in line 1 of Table 1).

## Appendix S1: Technical details.

All molecular geometries are taken from the GMTKN55 website.

Orbitals are computed with the Gaussian 16 program and a 200x590 grid. The orbitals are written to a WFN file for analysis by an in-house post-Gaussian program, "postG". We use postG to compute BR exchange, correlation, and dispersion energies only. All other energies, including the exact exchange energy, are taken from the Gaussian 16 output using keyword "Extralink=L608" to print the individual energy components to the output file.

Our postG computations employ 80 radial points on H and He atoms, 120 on atoms Li to Ne, 160 on atoms Na to Ar, 200 on atoms K to Kr, and 240 beyond Kr. We employ 302 angular points on all atoms. These are smaller grids than typically used in standard DFT computations (eg., 200x590 in our Gaussian 16 calculations) but the postG correlation and dispersion energies are a small fraction of the total energy.

PT2 energies are computed by the Q-Chem 6.0 program using "resolution of the identity" (RI-MP2) and frozen cores (as in Ref. 4 in the text). Single-excitation terms have been omitted. The Gaussian 16 and postG computations were run on facilities of the Digital Research Alliance of Canada. The PT2 computations were run on the "ChemFarm" HPC facility of the Weizmann Institute Faculty of Chemistry.

## Appendix S2: XDM dispersion and core pseudopotentials.

Throughout the initial "all-electron" GMTKN54 tests, all the density-functional terms and the XDM dispersion terms were exactly as given in Refs. 1 and 2 (Refs. 5 and 6 in the text). In our final calibration on the full GMTKN55 database, however, special consideration of the XDM dispersion terms was needed in order to accommodate core pseudopotentials.

The XDM dispersion coefficients for the atom pair $ij$ are given by

$$C_{6,ij} = \frac{\alpha_i \alpha_j <M_1^2>_i <M_1^2>_j}{<M_1^2>_i \alpha_j + <M_1^2>_j \alpha_i}$$

$$C_{8,ij} = \frac{3}{2} \frac{\alpha_i \alpha_j (<M_1^2>_i <M_2^2>_j + <M_2^2>_i <M_1^2>_j)}{<M_1^2>_i \alpha_j + <M_1^2>_j \alpha_i} \quad \text{(S1)}$$

$$C_{10,ij} = 2 \frac{\alpha_i \alpha_j (<M_1^2>_i <M_3^2>_j + <M_3^2>_i <M_1^2>_j)}{<M_1^2>_i \alpha_j + <M_1^2>_j \alpha_i} + \frac{21}{5} \frac{\alpha_i \alpha_j <M_2^2>_i <M_2^2>_j}{<M_1^2>_i \alpha_j + <M_1^2>_j \alpha_i}$$

with $\alpha_i$ the effective atom-in-molecule polarizability of atom $i$. We estimate this effective polarizability from the well-known proportionality between polarizability and volume. Thus, for $\alpha_i$, we use

$$\alpha_i = \left( \frac{<r^3>_i}{<r^3>_{i,free}} \right) \alpha_{i,free} = \left( \frac{\int r^3 w_i(\mathbf{r}) \rho(\mathbf{r}) d^3\mathbf{r}}{\int r^3 \rho_{i,free}(\mathbf{r}) d^3\mathbf{r}} \right) \alpha_{i,free} \quad \text{(S2)}$$

where $\alpha_{i,free}$ is the free atomic polarizability and the expectation value $<r^3>_i$ is assumed to be a measure of effective volume. The functions $w_i(\mathbf{r})$ are Hirshfeld atomic partitioning weights[3] defined by

$$w_i(\mathbf{r}) = \frac{\rho_i^{at}(\mathbf{r})}{\sum_n \rho_n^{at}(\mathbf{r})} \tag{S3}$$

where $\rho_i^{at}$ is the spin-depolarized and sphericalized free atomic density on center $i$ and the $n$ summation is over all atoms. Free atomic polarizabilities $\alpha_{i,free}$ and free atomic volumes $<r^3>_{i,free}$ are conveniently tabulated in Ref. 4.

The quantities $<M_\ell^2>_i$ are atomic expectation values of squared multipoles:

$$<M_\ell^2>_i = \sum_\sigma \int w_i(\mathbf{r}) \rho_\sigma(\mathbf{r}) \left[ r_i^\ell - (r_i - d_{X\sigma})^\ell \right]^2 d^3\mathbf{r} \tag{S4}$$

where $r_i$ is the distance from nucleus $i$, and $d_{X\sigma}$ is the magnitude of the dipole moment of the $\sigma$-spin exchange hole plus its reference electron at point $\mathbf{r}$. This dipole moment magnitude is approximated by the displacement $b$ in the Becke-Roussel model:

$$d_{X\sigma}(\mathbf{r}) = b \tag{S5}$$

(see Appendix A of Ref. 1). Note that $b$ is capped at $r_i$ in the above numerical integrations.

In Appendix E of Ref. 1, three dispersion damping schemes were discussed: "Becke-Johnson" (BJ) damping, an alternative energy-based damping, and an average of both. For our calibration on the full GMTKN55 database, and in our work going forward, we simplify to the energy-based damping scheme alone (involving only one damping parameter as opposed to two for BJ damping). The total dispersion energy is given by

$$E_{disp}^{XDM} = -\sum_{j>i} \left( \frac{C_{6,ij}}{(3\kappa C_{6,ij}/E_{C,ij}) + R_{ij}^6} + \frac{C_{8,ij}}{(3\kappa C_{8,ij}/E_{C,ij}) + R_{ij}^8} + \frac{C_{10,ij}}{(3\kappa C_{10,ij}/E_{C,ij}) + R_{ij}^{10}} \right), \tag{S6}$$

i.e., the damped contribution from the pair $ij$ at $R_{ij} = 0$ is $-E_{C,ij}/\kappa$, where $E_{C,ij}$ is the sum of the atomic correlation energies (absolute value) of atoms $i$ and $j$. Each of the $C_{6,ij}$, $C_{8,ij}$, and $C_{10,ij}$ terms is assumed to contribute an equal third to this limit. Previously, the atomic correlation energies were computed "on the fly" by a Hirshfeld partitioning of the opposite-spins-only LDA:[5]

$$E_{Ci} = \int w_i \left[ e_C^{LDA}(\rho/2, \rho/2) - 2e_C^{LDA}(\rho/2, 0) \right] d^3r \quad , \quad E_{C,ij} = |E_{Ci} + E_{Cj}| \tag{S7}$$

where $w_i$ is the Hirshfeld weight of Eq. (S3) and $e_C^{LDA}(\rho_\alpha, \rho_\beta)$ is the electron gas correlation energy per unit volume.[6] The opposite-spins-only LDA fortuitously gives rather good atomic correlation energies (see the column labelled "Stoll" in Table 3 in Ref. 7).

In core pseudopotential calculations, the "on the fly" method of Eq. (S7) will not work, as the valence pseudo-density will give erroneously low $|E_{Ci}|$. Going forward, then, we precompute $|E_{Ci}|$ for *free* atoms using spin-depolarized and sphericalized LDA densities. The results are tabulated below for main-group atoms, in atomic units:

| Atom | $|E_{Ci}|$ | Atom | $|E_{Ci}|$ | Atom | $|E_{Ci}|$ | Atom | $|E_{Ci}|$ | Atom | $|E_{Ci}|$ |
|---|---|---|---|---|---|---|---|---|---|
| H  | 0.0204 | Na | 0.415 | Ga | 1.41 | Sb | 2.48 | At | 4.54 |
| He | 0.0573 | Mg | 0.456 | Ge | 1.46 | Te | 2.54 | Rn | 4.59 |
| Li | 0.0825 | Al | 0.496 | As | 1.51 | I  | 2.59 |    |      |
| Be | 0.115  | Si | 0.539 | Se | 1.56 | Xe | 2.65 |    |      |
| B  | 0.151  | P  | 0.584 | Br | 1.62 | Cs | 2.69 |    |      |
| C  | 0.191  | S  | 0.631 | Kr | 1.67 | Ba | 2.74 |    |      |
| N  | 0.235  | Cl | 0.680 | Rb | 1.72 | Tl | 4.31 |    |      |
| O  | 0.281  | Ar | 0.730 | Sr | 1.76 | Pb | 4.36 |    |      |
| F  | 0.329  | K  | 0.767 | In | 2.38 | Bi | 4.42 |    |      |
| Ne | 0.380  | Ca | 0.808 | Sn | 2.43 | Po | 4.48 |    |      |

With only one damping parameter $\kappa$ in Eq. (S6), its value is easily determined by minimizing the WTMAD2(54). We obtain $\kappa = 850$, with B1LYP orbitals and the def2-QZVPP basis set.

Additional XDM-related issues need to be considered in pseudopotential calculations. How does the absence of core density affect the volume integral of Eq. (S2) and the moment integrals of Eq. (S4)? This has been investigated, again in *free* atoms, using spin-depolarized, sphericalized, relativistic[8] LDA densities. We find that relative core contributions are largest for the dipole ( $\ell = 1$ ) integral of Eq. (S4), as can be guessed by the powers of $r$ in the integrands of Eqs. (S2) and (S4). The def2 dipole core contributions are about 6% for Pb and Bi atoms, and about 3% for Sn, Sb, Te, and I atoms. We feel it safe to ignore these in the present work. As a check, full GMTKN55 computations have been performed with and without core corrections. The fits return slightly different coefficients, but the WTMAD2s are within 0.01 kcal/mol of each other. Elements further to the left in the periodic table will have relatively larger core corrections. Future work will establish if, or if not, they are generally necessary.

Appendix S3: Preliminary tests on 3d transition-metal reactions.

Transition-metal (TM) chemistry poses challenges to electronic structure methods beyond main-group and organic chemistry, due to lower orbital energy gaps and high potential for multi-reference character. Reliable benchmark sets for realistic TM reactions are few, but their number is slowly growing. Among them are the MOR41 (closed-shell metal organic reactions) database,[1] the MOBH35 (metal organic barrier heights) database,[2] and ROST61[3] (realistic open-shell transition-metal reactions). Full assessment of DH23 on these databases will be undertaken in upcoming work. Meanwhile, we have results of preliminary tests on the 3d reactions in MOR41.

MOR41 comprises 10 reactions of 3d transition metals, and 31 of 4d/5d metals. 3d reactions tend to be more problematic than 4d/5d reactions. The MOR41 paper thus includes separate analysis of each subset.[1] Here we compare early DH23 results with the MOR41 3d-subset statistics. For DH23 fit to GMTKN54, we obtain a 3d-subset MAD of 4.3 kcal/mol. This is lower than the 4.6 kcal/mol *average* MAD of the 17 double-hybrids tested in MOR41[1] (see Table S13 of the Supporting Information). We note that the *best* DH in the MOR41 paper, indeed the best functional overall, with a 3d-subset MAD of 1.9 kcal/mol, was PWPB95-D3(BJ).[4] This DH has *no* parallel-spin PT2. Parallel-spin PT2 appears to be detrimental in TM reaction computations.[1,5] We have therefore tested an "SOS"DH23 variant ("scaled opposite spin") with the parallel-spin PT2 term removed. This 11-parameter "SOS"DH23, fit to GMTKN54, has a much improved 3d-subset MAD of 2.6 kcal/mol. This is not far from the 1.9 kcal/mol of the best-performing PWPB95-D3(BJ) in MOR41[1], and equal to the MAD of the best-performing hybrid functional, ωB97X-V.[6] The WTMAD2(54) of "SOS"DH23 is only 0.14 kcal/mol higher than for DH23.

Complete details will be reported in a future paper.